\documentclass[11pt]{article}
\usepackage{setspace}
\usepackage{amssymb}
\usepackage{amsmath}
\usepackage{amsfonts}

\def\be{\begin{equation}}

\def\ee{\end{equation}}

\newcommand\vecbf[1]{{\bf #1}}
\newcommand\mat[1]{ {\mathbb #1}}
\newcommand\rqqq{\mat R}
\def\dd{\partial}

\newcommand\of[1]{\left( #1 \right)}
\newcommand\sqof[1]{\left[ #1 \right]}

\def\bea{\begin{eqnarray}}

\def\eea{\end{eqnarray}}

\def\half{\frac{1}{2}}

\newcommand\eps{\epsilon}

\begin{document}

\singlespace

\begin{flushright} BRX TH-635 \\
CALT 68-2825
\end{flushright}

\vspace*{.3in}

\begin{center}

{\Large\bf No Bel-Robinson Tensor for Quadratic Curvature Theories}

{\large S.\ Deser}

{\it Physics Department,  Brandeis University, Waltham, MA 02454 and \\
Lauritsen Laboratory, California Institute of Technology, Pasadena, CA 91125 \\
{\tt deser@brandeis.edu}
}

{\large J.\ Franklin}

{\it  Reed College, Portland, OR 97202 \\
{\tt jfrankli@reed.edu}}

\end{center}

\begin{abstract}
We attempt to generalize the familiar covariantly conserved Bel-Robinson tensor $B_{\mu\nu\alpha\beta}\sim R R$ of GR and its recent topologically massive third derivative order counterpart $B \sim R \, DR$, to quadratic curvature actions. Two very different models of current interest are examined: fourth order $D=3$ ``new massive", and second order $D>4$ Lanczos-Lovelock, gravity. On dimensional grounds, the candidates here become $B\sim DRDR+RRR$. For the $D=3$ model, there indeed exist conserved $B\sim \dd R \dd R$ in the linearized limit.   However, despite a plethora of available cubic terms, $B$ cannot be extended to the full theory. The $D>4$ models are not even linearizable about flat space, since their field equations are quadratic in curvature; they also have no viable $B$, a fact that persists even if one includes cosmological or Einstein terms to allow linearization about the resulting dS vacua. These results are an unexpected, if hardly unique, example of linearization instability.
\end{abstract}

\section{Introduction}

The Bel-Robinson tensor $B_{\mu\nu\alpha\beta}$ was first discovered in ordinary Einstein gravity (GR) at $D=4$, in a search for a gravitational counterpart of the usual matter stress-tensor $T_{\mu\nu}$. Since there can be neither local tensorial gravitational candidates of second derivative order (because quantities $\sim \dd g_{\mu\nu}\,  \dd g_{\alpha\beta}$, being frame-dependent, can be made to vanish at any point), nor any non-covariantly conserved ones, the successful candidate was indeed quadratic in the tensorial ``field strengths" -- curvatures $B \sim R R$, and covariantly conserved, in analogy with the quadratic Maxwell $T_{\mu\nu}\sim F\, F$.   Subsequently, conserved $B$ were found for arbitrary dimension and matter analogs have also been constructed (see, e.g., [1] for earlier references).

Very recently, we showed that a conserved $B\sim R \, DR$ could be defined for topologically massive gravity [2].  Given this theory's third derivative order, we speculated there on extending $B$  to other gravitational systems, in particular to the currently popular quadratic curvature models.  The present note reports a negative outcome in two active, very different theories. For the fourth derivative $D=3$ new massive gravity (NMG) [3], without an Einstein term [4] for simplicity, there is a $B\sim \dd R \dd R$ at linearized level but it cannot be extended to the full theory, despite an enormous number of possible cubic correction terms. The Lanczos-Lovelock (LL) [5] models' quadratic curvature field equations, $\sim RR=0$, obviously cannot even be linearized about flat space and turn out to have no conserved $B$ either. Adding cosmological or Einstein terms to permit linearization (to effective cosmological GR) about the resulting dS vacua only allows the usual linear $B\sim RR$ of GR, but no nonlinear extension.

\section{Fourth Order Models}

Since the machinery is considerably more complicated here than in GR, we
analyze the (purely quadratic part of) NMG [3]:
\begin{equation}
I = \int d^3 \, x \, \sqrt{-g} \, \of{\bar R^{\mu\nu}\, \bar R_{\mu\nu} - \bar R^2},   \, \, \, \, \, \, \, \, \, \, \, \, 
\bar R_{\mu\nu} \equiv R_{\mu\nu} - \frac{1}{4} \, g_{\mu\nu} \, R.
\end{equation}
[The Cotton tensor of TMG is the curl of the Schouten tensor $\bar R_{\mu\nu}$.]  The resulting field equations, 
\begin{equation}
\Box \bar R_{\mu\nu} - D_\mu D_\nu \, \bar R - 4\, \bar R_\mu^{\, \, \, \sigma} \, \bar R_{\nu \sigma} +  \bar R \, \bar R_{\mu\nu} +  g_{\mu\nu} \, \bar R^2 = 0
\end{equation}
of course obey Bianchi identities since (1) is an invariant.    The field equations begin as $DD\bar R\sim 0$, so $B$ cannot just imitate the $RR$ of GR, but must instead start as $B\sim D\bar R \, D\bar R$, along with
possible cubic, $\bar R \bar R\bar R$, terms since they have the same dimension as $D\bar R\, D\bar R$. Schematically then, we expect that
\begin{equation}
B \sim D\bar R \, D\bar R + \bar R \bar R \bar R.
\end{equation}
[One can also consider similar terms $\sim D(\bar R D\bar R)$ or $\bar RDD\bar R$, but they are already irrelevant at linear level.]  We begin with the linearized (about flat space) truncation, $\dd \dd \bar R=0$, of (2); the corresponding linearized $B$ is simply given by
\begin{equation}
B_{\mu\nu\alpha\beta} =  \bar R_{\alpha\beta,}^{\, \, \, \, \, \, \, \,  \, \sigma}\, \of{ \bar R_{\sigma \nu,\mu} - \bar R_{\mu\nu,\sigma} }.
\end{equation}
Its conservation,
\begin{equation}
\dd^\mu \, B_{\mu\nu\alpha\beta} = 0
\end{equation}
is verified using the field equations, the Bianchi identity, and antisymmetry of the parenthesis in (4). Conservation also holds for all permutations of $B$'s $(\nu \alpha \beta)$ indices; for completeness, we note also the identically conserved, if irrelevant, $b_{\mu\nu\alpha\beta} = D^\gamma \, H_{[\gamma\mu] \nu\alpha\beta}$, $H$ antisymmetric in $\gamma \leftrightarrow \mu$.

Thus encouraged, we turn to candidates (3) at nonlinear level, where
the now covariant derivatives $D_\mu$ no longer commute. At first sight, there are so many available independent terms cubic in $\bar R_{\mu\nu}$ and $g_{\mu\nu}\, \bar R$, that success seems guaranteed; however, it turns out to be unattainable.  The simplest procedure is to take the covariant divergence of the
initial $B\sim D\bar R\, (D\bar R-D\bar R)$ ansatz and  try to compensate for the resulting cubic $\sim \bar R\bar R\, D\bar R$ terms by adding suitable $\bar R\bar R\bar R$ to $B$. The procedure is straightforward, using the $D=3$ properties 
\begin{equation}
\begin{aligned}
R_{\mu\nu\rho\sigma} &= g_{\mu\rho} \, \bar R_{\nu\sigma} + g_{\nu\sigma} \, \bar R_{\mu\rho} - g_{\nu\rho} \, \bar R_{\mu\sigma} - g_{\mu\sigma} \, \bar R_{\nu\rho}, \\
[D_\alpha, D_\beta] \, V^\gamma &= \sqof{\delta^\gamma_\beta \, \bar R_{\sigma\alpha}  - \delta^\gamma_\alpha \, \bar R_{\sigma\beta} + g_{\sigma\alpha} \, \bar R^\gamma_{\, \, \beta} - g_{\sigma\beta} \, \bar R^\gamma_{\, \, \alpha}}\, V^\sigma. 
\end{aligned}
\end{equation}
We find 
\begin{equation}
D^\mu \, B_{\mu\nu\alpha\beta} = \sqof{ \bar R^{\rho\mu} \, \bar R_{\rho\beta} \, \of{\bar R_{\mu\nu;\alpha} - \bar R_{\alpha\nu;\mu} } + \half\,  \bar R_{\alpha\beta;\sigma} \, \of{\bar R^{\sigma\rho} \, \bar R_{\rho\nu} - \bar R^\sigma_{\, \, \nu} \, \bar R} }+ \left( \alpha \leftrightarrow \beta\right),
\end{equation}
to which could be added the above $(\alpha \beta \nu)$ permutations. The problem is now to find, among all $(\bar R\, \bar R\, \bar R)_{\mu\nu\alpha\beta}$, those whose $\mu$-divergence cancels (7), or the sum of permutations.  Note that (7) contains both $\mu$- and $\alpha\, , \, \beta$-derivatives, and that the latter must arise from cubic terms of the form $\sim g_{\mu\alpha} (\bar R\, \bar R\, \bar R)_{\nu\beta}$, and possibly $g_{\mu\alpha}\, g_{\nu\beta} \, (\bar R \, \bar R \, \bar R)$; these are also plentiful.  This systematic approach failed: any compensating term also created a new problematic one. We then undertook the brute force approach, involving the sum of {\it all} dimensionally possible candidate terms, $\sim D\bar R\,  D\bar R+\bar R\, DD\bar R+ \bar R \bar R \bar R$; to be sure, any $D\bar R \, D\bar R$ combinations that fail the linearized conservation test could be excluded {\it a priori}.

We sketch the procedure and counts, but omit the uninstructive, laborious, details\footnote{These may be found in the notebook at: \texttt{http://people.reed.edu/$\sim$jfrankli/NullSpace.nb}.}.  Start with the thirty nine possible terms of the form $\bar R\bar R\bar R$, incuding $g\, \bar R \bar R \bar R$ terms, together with the $75$ $D (\bar R\,  D\bar R)$ candidates, giving a total of $m = 114$ potential terms.  Take the divergence of each of these, enforcing the on-shell, Bianchi and three-dimensional simplifications. That gives, for each term, a right-hand side that is  a linear combination of $(DD \bar R) \, D\bar R$, $\bar R \bar R D\bar R $ and $\bar R \, DDD\bar R$.  There are $n = 231$ unique terms of this form in these sums.  
We can view the action of $D^\mu$ as a linear operator taking the $m$ starting ingredients to a space of $n$ outputs:  assign to each candidate term a basis vector $\{\vecbf e_i\}_{i=1}^{m}$, and to each term appearing in the sums $D^\mu \, \vecbf e_i$, the basis vector $\{\vecbf E_i\}_{i=1}^{n}$, the matrix $\mat D \in \rqqq^{n\times m}$ just maps $\vecbf e_k$ to its sum: $\mat D \, \vecbf e_k = \sum_{i=1}^{n} \alpha_i \, \vecbf E_i$.  We constructed this matrix using a symbolic algebra package, and found that $\mat D$ has no null space, and so no conserved Bel-Robinson tensor exists.

\section{Second Order, LL, Theories}

At the other end of the quadratic curvature spectrum from the $D=3$ model (1) lie the $D> 4$ LL theories, whose field equations remain of second derivative order, but do not admit any linearized expansion about flat space. The Lagrangians are, in vielbein formulation,
\begin{equation}\label{LGBL}
L=\eps^{abcdf\ldots}\,  \eps^{ABCDF\ldots}\,  R_{abAB} \, R_{cdCD}\,  e_{fF} \ldots  
\end{equation}
The field equations simply consist of removing one vielbein factor,
\begin{equation}\label{FEGBL} 
E^{fF} \equiv (\eps \eps\,  RR\, e)^{fF}=0,
\end{equation}
since the curvatures' variations vanish by the cyclic identities.   Lowercase indices are local, capitals are world. Diffeomorphism invariance of the action ensures that  $D_F\,  E^{fF}$ vanishes identically, by the cyclic identities. Indeed, this is even true in $D>5$: if we open any second pair of indices $(gG)$, there results an identically conserved, hence uninteresting, $H^{fg\ldots FG \ldots}$.  We cannot even start with a linearized $B$, since the field equations (9) are intrinsically nonlinear. That there is no true $B$ is rather clear: for example, attaching a curvature to $H$ above loses conservation, which no additional cubics can help, while no $DR \, DR$ term is relevant just because it is  linearizable.  Extending LL to include a cosmological or Einstein term does allow for linearization about a nonflat, dS, vacuum: a dS metric ansatz reduces (9) to the equation $\sim a\, \Lambda^2=b \, \Lambda$ in either choice. There is always a flat, $\Lambda=0$, but also now a dS, $\Lambda=b/a$, vacuum\footnote{While the appearance of a second, dS, vacuum in $R+$LL -- indeed in all $R+R^2$ models but ($D=4$) $R+$Weyl$^2$ -- 
 is reasonably clear [6], it may seem counter-intuitive that, unlike GR, quadratic models with cosmological term allow both flat and dS vacua.}. About the latter, the
system linearizes to cosmological Einstein, which of course enjoys the same conserved $B\sim RR$ as $\Lambda=0$ GR. 
However, just like its predecessor in NMG, this $B$ evaporates as soon as we go to the full theory, since there
is still no way to take into account the nonlinear $RR$ part of (9).

\section{Conclusion}

We conclude that both NMG and Einstein$+$LL violate Fermi's rule: if the lowest order works, the conjecture is correct, while pure LL doesn't even have a lowest (here linearized) order. 
Clearly there is no invariance here like that [1] underlying $B$-conservation in GR.

\noindent SD acknowledges support from NSF PHY-1064302 and DOE DE-FG02-164 92ER40701 grants.

\end{document}